\documentclass[aps,twocolumn,prl,floatfix,showpacs]{revtex4-1}

\usepackage{amsmath}
\usepackage{amssymb}
\usepackage{amsfonts}
\usepackage[pdftex]{graphicx}
\usepackage{units}
\usepackage{bm}
\usepackage{color}
\usepackage{tikz}
\usepackage{array}
\usepackage{hyperref}
\hypersetup{colorlinks=true,allcolors=blue}

\definecolor{dunkelgrau}{rgb}{0.8,0.8,0.8}
\definecolor{hellgrau}{rgb}{0.95,0.95,0.95}
\newcommand{\cone}{\mathrm{i}}
\newcommand{\operator}[1]{\ensuremath{\hat{#1}}}
\newcommand{\mat}[1]{\boldsymbol{\mathsf{#1}}}

\renewcommand{\vec}[1]{\boldsymbol{#1}}
\newcommand{\bra}[1]{\langle {#1} |}
\newcommand{\ket}[1]{| {#1} \rangle}

\begin{document}
\title{Magnetic moment of inertia within the breathing model}

\author{Danny Thonig}
\email{danny.thonig@physics.uu.se}
\affiliation{Department of Physics and Astronomy, Material Theory, University Uppsala, S-75120 Uppsala, Sweden}

\author{Manuel Pereiro}
\affiliation{Department of Physics and Astronomy, Material Theory, University Uppsala, S-75120 Uppsala, Sweden}

\author{Olle Eriksson}
\affiliation{Department of Physics and Astronomy, Material Theory, University Uppsala, S-75120 Uppsala, Sweden}

\date{\today}

\begin{abstract}
An essential property of magnetic devices is the relaxation rate in magnetic switching which strongly depends on the energy dissipation  and magnetic inertia of the magnetization dynamics. Both parameters are commonly taken as a phenomenological entities. However very recently, a large effort has been dedicated to obtain Gilbert damping from first principles.   In contrast, there is no {\it ab initio} study that so far has reproduced measured data of magnetic inertia in magnetic materials. In this letter, we present and elaborate on a theoretical model for calculating the magnetic moment of inertia based on the torque-torque correlation model. Particularly, the method has been applied to bulk bcc Fe, fcc Co and fcc Ni in the framework of the tight-binding approximation and the numerical values are comparable with recent experimental measurements. The theoretical results elucidate the physical origin of the moment of inertia based on the electronic structure. Even though the moment of inertia and damping are produced by the spin-orbit coupling, our analysis shows that they are caused by undergo different electronic structure mechanisms.

\end{abstract}









\maketitle

The research on magnetic materials with particular focus on spintronics or magnonic applications became more and more intensified, over the last decades \cite{Parkin03,Xu06}. For this purpose, ``good'' candidates are materials exhibiting thermally stable magnetic properties \cite{Miyamachi13}, energy efficient magnetization dynamics \cite{Kim16,Chumak15}, as well as fast and stable magnetic switching \cite{Tudosa04, Chudnovskiy14}. Especially the latter can be induced by \textit{i)} an external magnetic field, \textit{ii)} spin polarized currents \cite{Maekawa12}, \textit{iii)} laser induced all-optical switching \cite{Kimel14}, or \textit{iv)} electric fields \cite{Stoehr09}. The aforementioned magnetic excitation methods allow switching of the magnetic moment on sub-$\unit{ps}$ timescales. 

\begin{figure}
\centering
\includegraphics[width=0.6\columnwidth]{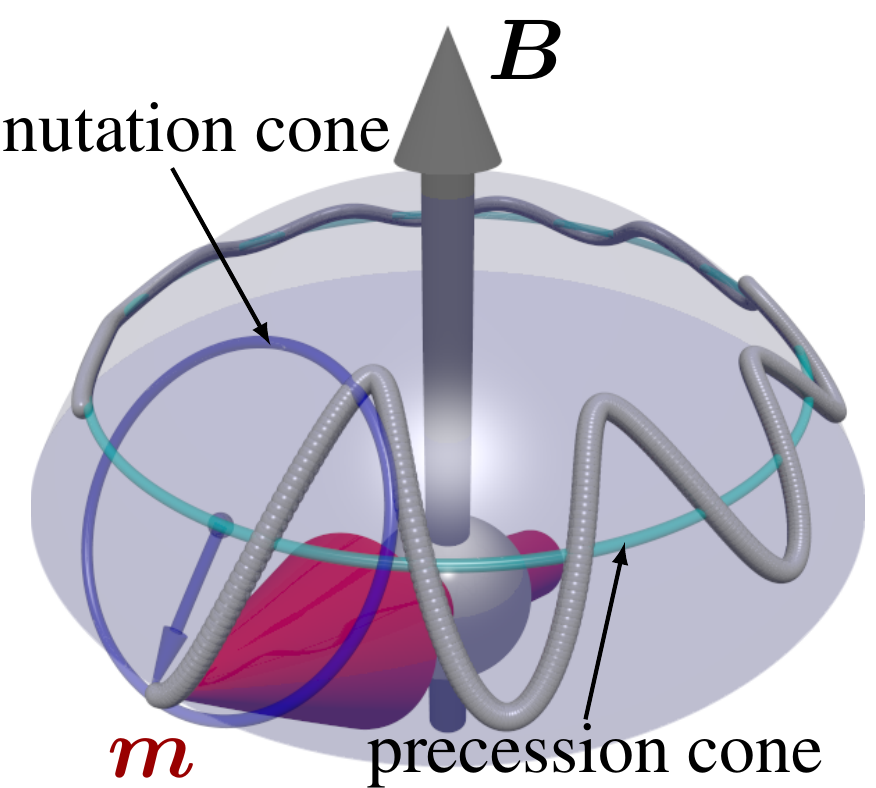}
\caption{(Color online) Schematic figure of nutation in the atomistic magnetic moment evolution. The magnetic moment $\vec{m}$ (red arrow) evolves around an effective magnetic field $\vec{B}$ (gray arrow) by a superposition of the precession around the field (bright blue line) and around the angular momentum axis (dark blue line). The resulting trajectory (gray line) shows an elongated cycloid.}
\label{fig:1}
\end{figure}

The classical atomistic Landau-Lifshitz-Gilbert (LLG) equation \cite{Antropov96,Skubic08} provides a proper description of magnetic moment switching \cite{Chimata15}, but is derived within the adiabatic limit \cite{Born28,Kato50}. This limit characterises the blurry boundary where the time scales of electrons and atomic magnetic moments are separable \cite{Moriya85} --- usually between $\unit[10-100]{fs}$. In this time-scale, the applicability of the atomistic LLG equation must be scrutinized in great detail. In particular, in its common formulation, it does not account for creation of magnetic inertia \cite{Boettcher12c}, compared to its classical mechanical counterpart of a gyroscope. At short times, the rotation axis of the gyroscope do not coincide with the angular momentum axis due to a ``fast'' external force. This results in a superimposed precession around the angular-momentum and the gravity field axis; the gyroscope nutates. It is expected for magnetisation dynamics that atomic magnetic moments behave in an analogous way on ultrafast timescales \cite{Ciornei10,Boettcher12c} (Fig.~\ref{fig:1}). 

Conceptional thoughts in terms of ``magnetic mass'' of domain walls were already introduced theoretically by D\"oring \cite{Doering48} in the late 50's and evidence was found experimentally by De Leeuw and Robertson \cite{DeLeeuw75}. More recently, nutation was discovered on a single-atom magnetic moment trajectory in a Josephson junction \cite{Zhu06,Fransson08,Fransson08b} due to angular momentum transfer caused by an electron spin flip. From micromagnetic Boltzman theory, Ciornei et al. \cite{Ciornei10,Ciornei11} derived a term in the extended LLG equation that addresses ``magnetic mass'' scaled by the moment of inertia tensor $\iota$. This macroscopic model was transferred to atomistic magnetization dynamics and applied to nanostructures by the authors of Ref.~\onlinecite{Boettcher12c}, and analyzed analytically in Ref.~\onlinecite{Olive15} and Ref.~\onlinecite{Kikuchi15}. Even in the dynamics of Skyrmions, magnetic inertia was observed experimentally \cite{Buettner15}. 

Like the Gilbert damping $\alpha$, the moment of inertia tensor $\iota$ have been considered as a parameter in theoretical investigations and postulated to be material specific. Recently, the latter was experimentally examined by Li et al. \cite{Li15b} who measured the moment of inertia for Ni$_{79}$Fe$_{21}$ and Co films near room temperature with ferromagnetic resonance (FMR) in the high-frequency regime (around $\unit[200]{GHz}$). At these high frequencies, an additional stiffening was observed that was quadratic in the probing frequency $\omega$ and, consequently, proportional to the moment of inertia $\iota=\pm\alpha\cdot\tau$. Here, the lifetime of the nutation $\tau$ was determined to be in the range of $\tau=\unit[0.12-0.47]{ps}$, depending not only on the selected material but also on its thickness. This result calls for a proper theoretical description and calculations based on {\it ab-initio} electronic structure footings.

A first model was already provided by Bhattacharjee et al. \cite{Bhattacharjee12}, where the moment of inertia $\iota$ was derived in terms of Green's functions in the framework of the linear response theory. However, neither first-principles electronic structure-based numerical values nor a detailed physical picture of the origin of the inertia and a potential coupling to the electronic structure was reported in this study. In this Letter, we derive a model for the moment of inertia tensor based on the torque-torque correlation formalism \cite{Kambersky84,Gilmore07b}. We reveal the basic electron mechanisms for observing magnetic inertia by calculating numerical values for bulk itinerant magnets Fe, Co, and Ni with both the torque-torque correlation model and the linear response Green's function model \cite{Bhattacharjee12}. Interestingly, our study elucidate also the misconception about the sign convention of the moment of inertia \cite{Faehnle13}. 

The moment of inertia $\iota$ is defined in a similar way as the Gilbert damping $\alpha$ within the effective dissipation field $B_{\mathrm{diss}}$ 
\cite{Kambersky70,Kambersky84}. This \textit{ad hoc} introduced field is expanded in terms of viscous damping $\alpha\nicefrac{\partial \vec{m}}{\partial t}$ and magnetic inertia $\iota\nicefrac{\partial^2 \vec{m}}{\partial t^2}$ in the relaxation time approach \cite{Faehnle11,Faehnle13} (see Supplementary Material). The off-equilibrium magnetic state induces excited states in the electronic structure due to spin-orbit coupling. Within the adiabatic limit, the electrons equilibrate into the ground state at certain time scales due to band transitions \cite{Gilmore07}. If this relaxation time $\tau$ is close to the adiabatic limit, it will have two implications for magnetism: \textit{i)} magnetic moments respond in a inert fashion, due to formation of magnetism, \textit{ii)} the kinetic energy is proportional to $\nicefrac{\mathrm{m} \vec{u}^2}{2}$ with the velocity $\vec{u}=\nicefrac{\partial \vec{m}}{\partial t}$ and the ``mass'' $\mathrm{m}$ of magnetic moments, following equations of motion of classical Newtonian mechanics. The inertia forces the magnetic moment to remain in their present state, represented in the Kambersky model by $\alpha=-\iota \cdot \tau$ (Ref.~\onlinecite{Faehnle11,Faehnle13}); the \textit{raison d'etre} of inertia is to behave opposite to the Gilbert damping.

In experiments, the Gilbert damping and the moment of inertia are measurable from the diagonal elements of the magnetic response function $\mat{\chi}$ via ferromagnetic resonance \cite{Gilmore07b} (see Supplementary Material)
\begin{align}
\alpha &= \frac{\omega_0^2}{\omega_M}\lim_{\omega\rightarrow 0}\frac{\Im\chi^\perp}{\omega}\\
\iota    &= \frac{1}{2}\frac{\omega_0^2}{\omega_M}\lim_{\omega\rightarrow 0}\frac{\partial_\omega\Re\chi^\perp}{\omega}-\frac{1}{\omega_0},
\label{eq:defiotaalpha}
\end{align}
\noindent where $\omega_M=\gamma B$ and $\omega_0=\gamma B_0$ are the frequencies related to the internal effective and the external magnetic field, respectively.  Thus, the moment of inertia $\iota$ is equal to the change of the FMR peak position, say the first derivative of the real part of $\chi$ with respect to the probing frequency \cite{Bhattacharjee12,Brataas08}. Alternatively, rapid external field changes induced by spin-polarized currents lead also to nutation of the macrospin \cite{Zhou10}. 

Setting $\chi$ on {\it ab-initio} footings, we use the torque-torque correlation model, as applied for the Gilbert damping in Ref.~\onlinecite{Kambersky84,Gilmore07}. We obtain (see Supplementary Material)
\begin{align}
\alpha^{\mu\nu}=\frac{g\pi}{m_s}\sum_{nm}\int T^\mu_{nm}\left(\vec{k}\right)T^\nu_{nm}\left(\vec{k}\right) W_{nm} \mathrm{d}\vec{k}\\
\iota^{\mu\nu}=-\frac{g\hbar}{m_s}\sum_{nm}\int T^\mu_{nm}\left(\vec{k}\right)T^\nu_{nm}\left(\vec{k}\right) V_{nm}\mathrm{d}\vec{k},
\label{eq:torque}
\end{align}
\noindent where $\mu,\nu=x,y,z$ and $m_s$ is the size of the magnetic moment. The spin-orbit-torque matrix elements $\vec{T}_{nm}=\bra{n,\vec{k}}\left[\vec{\sigma},H_{soc}\right]\ket{m,\vec{k}}$ --- related to the commutator of the Pauli matrices $\vec{\sigma}$ and the spin-orbit Hamiltonian --- create transitions between electron states $\ket{n,\vec{k}}$ and $\ket{m,\vec{k}}$ in bands $n$ and $m$. This mechanism is equal for both, Gilbert damping and moment of inertia. Note that the wave vector $\vec{k}$ is conserved, since we neglect non-uniform magnon creation with non-zero wave vector. The difference between moment of inertia and damping comes from different weighting mechanism $W_{nm},V_{nm}$: for the damping $W_{nm}=\int \eta(\varepsilon)A_{n\vec{k}}(\varepsilon)A_{m\vec{k}}(\varepsilon)\mathrm{d}\varepsilon$ where the electron spectral functions are represented by Lorentzian's $A_{n\vec{k}}(\varepsilon)$ centred around the band energies $\varepsilon_{n\vec{k}}$ and broadened by interactions with the lattice, electron-electron interactions or alloying. The width of the spectral function $\Gamma$  provides a phenomenological account for angular momentum transfer to other reservoirs. For inertia, however, $V_{nm}=\int f(\varepsilon)\left(A_{n\vec{k}}(\varepsilon)B_{m\vec{k}}(\varepsilon)+B_{n\vec{k}}(\varepsilon)A_{m\vec{k}}(\varepsilon)\right)\mathrm{d}\varepsilon$ where $B_{m\vec{k}}(\varepsilon)=\nicefrac{2(\varepsilon-\varepsilon_{m\vec{k}})\left((\varepsilon-\varepsilon_{m\vec{k}})^2 -3\Gamma^2\right)}{\left(\left(\varepsilon-\varepsilon_{m\vec{k}}\right)^2+\Gamma^2\right)^3}$ (see Supplementary Material). Here, $f(\varepsilon)$ and $\eta(\varepsilon)$ are the Fermi-Dirac distribution and the first derivative of it with respect to $\varepsilon$. Knowing the explicit form of $B_{m\vec{k}}$, we can reveal particular properties of the moment of inertia: \textit{i)} for $\Gamma\rightarrow 0$ ($\tau\rightarrow\infty$), $V_{nm}= \nicefrac{2}{\left(\varepsilon_{n\vec{k}}-\varepsilon_{m\vec{k}}\right)^3}$. Since $n=m$ is not excluded, $\iota \rightarrow-\infty$; the perturbed electron system will not relax back into the equilibrium. \textit{ii)} In the limit $\Gamma\rightarrow \infty$ ($\tau\rightarrow 0$), the electron system equilibrates immediately into the ground state and, consequently, $\iota=0$. These limiting properties are consistent with the expression $\iota=-\alpha\cdot\tau$. Eq.~\eqref{eq:torque} also indicates that the time scale is dictated by $\hbar$ and, consequently, on a femto-second time scale.

To study these properties, we performed first-principles tight binding (TB) calculations \footnote{\textit{CAHMD}, 2013. A computer program package for atomistic magnetisation dynamics simulations. Available from the authors; electronic address: danny.thonig@physics.uu.se.} of the torque-correlation model as described by Eq.~\eqref{eq:torque} as well as for the Green's function model reported in Ref.~\onlinecite{Bhattacharjee12}. The materials investigated in this letter are bcc Fe, fcc Co, and fcc Ni. Since our magnetic moment is fixed in the $z$ direction, variations occur primarily in $x$ or $y$ and, consequently, the effective torque matrix element is $T^- =\bra{n,\vec{k}}\left[\sigma^-,H_{soc}\right]\ket{m,\vec{k}}$, where $\sigma^- =\sigma_x - \cone \sigma_y$. The cubic symmetry of the selected materials allows only diagonal elements in both damping and moment of inertia tensor. 
\begin{figure}
\centering
\includegraphics [width = 0.85\columnwidth] {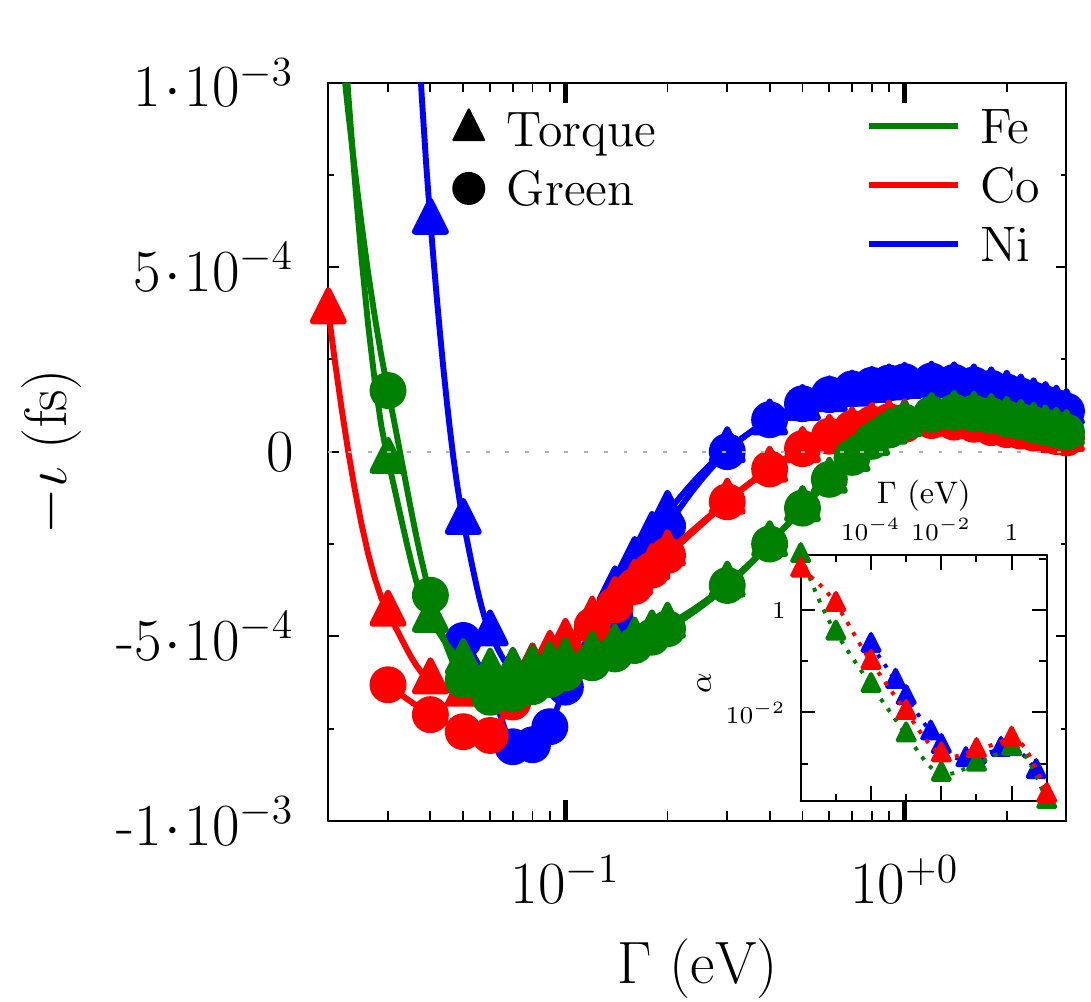}
\caption{(Color) Moment of inertia $\iota$ as a function of the band width $\Gamma$ for bcc Fe (green dotes and lines), fcc Co (red dotes and lines), and fcc Ni (blue dotes and lines) and with two different methods: \textit{i)} the torque-correlation method (filled triangles) and the \textit{ii)} Greens function method \cite{Bhattacharjee12}(filled circles). The dotted gray lines indicating the zero level. The insets show the calculated Gilbert damping $\alpha$ as a function of $\Gamma$. Lines are added to guide the eye. Notice the negative sign of the moment of inertia.}
\label{fig2}
\end{figure}
The numerical calculations, as shown in Fig.~\ref{fig2}, give results that are consistent with the torque-torque correlation model predictions in both limits, $\Gamma \rightarrow 0$ and $\Gamma \rightarrow \infty$. Note that the latter is only true if we assume the validity of the adiabatic limit up to $\tau=0$. It should also be noted that Eq.~\eqref{eq:torque} is only valid in the adiabatic limit ($>\unit[10]{fs}$). The strong dependency on $\Gamma$ indicates, however, that the current model is not a parameter-free approach. Fortunately, the relevant  parameters can be extracted from {\it ab-initio} methods: e.g., $\Gamma$ is related either to the electron-phonon self energy \cite{Pavarini13} or to electron correlations \cite{Sayad16}. 

The approximation $\iota=-\alpha\cdot\tau$ derived by F\"ahnle et al. \cite{Faehnle13} from the Kambersk\'y model is not valid for all $\Gamma$. It holds for $\Gamma<10$ meV, where intraband transitions dominate for both damping and moment of inertia; bands with different energies narrowly overlap. Here, the moment of inertia decreases proportional to $\nicefrac{1}{\Gamma^4}$ up to a certain minimum. Above the minimum and with an appropriate large band width $\Gamma$, interband transitions happen so that the moment of inertia approaches zero for high values of $\Gamma$. In this range, the relation $\iota=\alpha\cdot\tau$ used by Ciornei et al \cite{Ciornei10} holds and softens the FMR resonance frequency. Comparing qualitative the difference between the itinerant magnets Fe, Co and Ni, we obtain similar features in $\iota$ and $\alpha$ vs. $\Gamma$, but the position of the minimum and the slope in the intraband region varies with the elements: $\iota_{\mathrm{min}}=\unit[5.9\cdot 10^{-3}]{fs^{-1}}$ at $\Gamma=\unit[60]{meV}$ for bcc Fe,  $\iota_{\mathrm{min}}=\unit[6.5 \cdot 10^{-3}]{fs^{-1}}$ at $\Gamma=\unit[50]{meV}$ for fcc Co, and $\iota_{\mathrm{min}}=\unit[6.1\cdot 10^{-3}]{fs^{-1}}$ at $\Gamma=\unit[80]{meV}$ for fcc Ni. The crossing point of intra- and interband transitions for the damping was already reported by Gilmore et al. \cite{Gilmore07} and Thonig et al. \cite{Thonig13b}. The same trends are also reproduced by applying the Green's function formalism from Bhattacharjee et al. \cite{Bhattacharjee12} (see Fig.~\ref{fig2}). Consequently, both methods --- torque-torque correlation and the linear response Green's function method --- are equivalent as it can also be demonstrated not only for the moment of inertia but also for the Gilbert damping $\alpha$ (see Supplementary Material)\cite{Thonig13b}. In the torque-torque correlation model \eqref{eq:torque}, the coupling $\Gamma$ defines the width of the energy window in which transitions  $T_{nm}$ take place. The Green function approach, however, provides a more accurate description with respect to the {\it ab initio} results than the torque-torque correlation approach. This may be understood from the fact that a finite $\Gamma$ broadens and slightly shifts maxima in the spectral function. In particular, shifted electronic states at energies around the Fermi level causes differences in the minimum of $\iota$ in both models. Furthermore, the moment of inertia can be resolved by an orbital decomposition and, like the Gilbert damping $\alpha$, scales quadratically with the spin-orbit coupling $\zeta$, caused by the torque operator $\operator{T}$ in Eq.~\eqref{eq:torque}. Thus, one criteria for finding large moments of inertia is by having materials with strong spin-orbit coupling.
\begin{figure}
\centering
\includegraphics [width = 0.9\columnwidth] {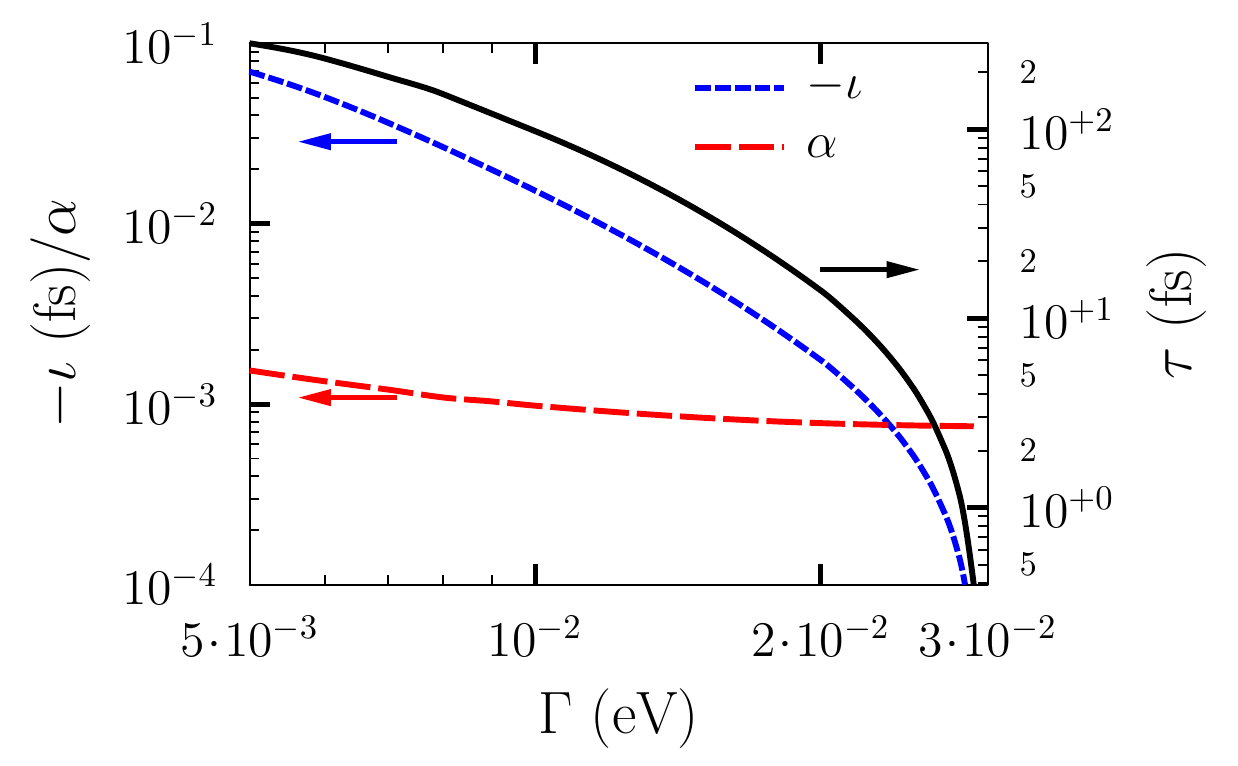}
\caption{(Color online) Gilbert damping $\alpha$ (red dashed line), moment of inertia $\iota$ (blue dashed line), and the resulting nutation lifetime $\tau=\nicefrac{\iota}{\alpha}$ (black line) as a function of $\Gamma$ in the intraband region for Fe bulk. Arrows indicating the ordinate belonging of the data lines. Notice the negative sign of the moment of inertia.}
\label{fig3}
\end{figure}

In order to show the region of $\Gamma$ where the approximation $\iota=-\alpha\cdot\tau$ holds, we show in Fig. \ref{fig3} calculated values of  $\iota$, $\alpha$, and the resulting nutation lifetime $\tau$ for a selection of $\Gamma$ that are below $\iota_{\mathrm{min}}$. According to the data reported in Ref.~\onlinecite{Li15b}, this is a suitable regime accessible for experiments. To achieve the room temperature measured experimental values of  $\tau=\unit[0.12-0.47]{ps}$, we have furthermore to guarantee that $\iota >> \alpha$. An appropriate experimental range is  $\Gamma\approx\unit[5-10]{meV}$, which is realistic and caused, e.g., by the electron-phonon coupling. A nutation lifetime of $\tau\approx\unit[0.25-0.1]{ps}$ is revealed for these values of $\Gamma$ (see Fig.~\ref{fig3}), a value similar to that found in experiment. The aforementioned electron-phonon coupling, however, is underestimated compared to the electron-phonon coupling from a Debye model ($\Gamma\approx\unit[50]{meV}$) \cite{Huefner07}. In addition, effects on spin disorder and electron correlation are neglected, that could lead to uncertainties in $\Gamma$ and hence discrepancies to the experiment. On the other hand, it is not excluded that other second order energy dissipation terms, $B_{\mathrm{diss}}$, proportional to $(\nicefrac{\partial \vec{e}}{\partial t})^2$ will also contribute \cite{Faehnle13} (see Supplementary material). The derivation of the moment of inertia tensor from the Kambersk\'y model and our numerics corroborates that recently observed properties of the Gilbert damping will be also valid for the moment of inertia: \textit{i)} the moment of inertia is temperature dependent \cite{Ebert11, Thonig13b} and decays with increasing phonon temperature, where the later usually increase the electron-phonon coupling $\Gamma$ in certain temperature intervals \cite{Huefner07}; \textit{ii)} the moment of inertia is a tensor, however, off-diagonal elements for bulk materials are negligible small; \textit{iii)} it is non-local \cite{Brataas08,Gilmore09,Thonig13b} and depends on the magnetic moment \cite{Steiauf05,Faehnle06,Yuan14}. Note that the sign change of the moment of inertia also effects the dynamics of the magnetic moments (see Supplementary Material).
\begin{figure}
\centering
\includegraphics [width = 0.75\columnwidth] {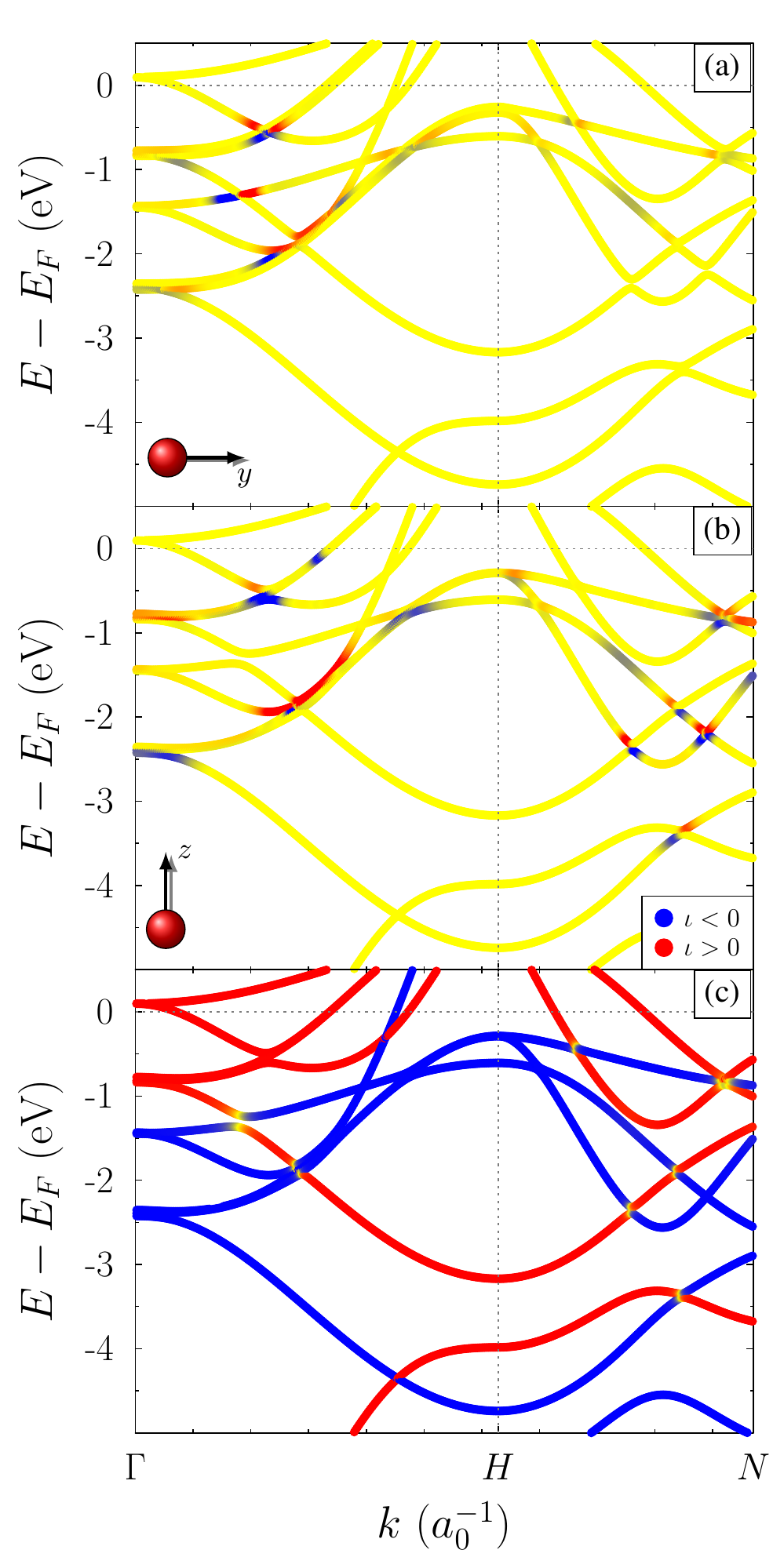}
\caption{(Color online) Moment of inertia in the electron band structure for bulk fcc Co with the magnetic moment a) in $y$ direction and b) in $z$ direction. The color and the intensity indicates the sign and value of the inertia contribution (blue - $\iota<0$; red - $\iota > 0$; yellow - $\iota \approx 0$). The dotted gray line is the Fermi energy and $\Gamma$ is $\unit[0.1]{eV}$. c) Spin polarization of the electronic band structure (blue - spin down; red - spin up; yellow - mixed states).}
\label{fig4}
\end{figure}

The physical mechanism of magnetic moment of inertia becomes understandable from an inspection of the electron band structure (see Fig.~\ref{fig4} for fcc Co, as an example). The model proposed here allows to reveal the inertia $\vec{k}$- and band-index $n$ resolved contributions (integrand of Eq.~\eqref{eq:torque}). Note that we analyse for simplicity and clarity only one contribution, $A_{n}B_{m}$, in the expression for $V_{nm}$. As Fig.~\ref{fig4} shows the contribution to $V_{nm}$ is significant only for specific energy levels and specific k-points. The figure also shows a considerable anisotropy, in the sense that magnetisations aligned along the z- or y-directions give significantly different contributions. Also, a closer inspection shows that degenerate  or even close energy levels $n$ and $m$, which overlap due to the broadening of energy levels, e.g. as caused by electron-phonon coupling, $\Gamma$, accelerate the relaxation of the electron-hole pairs caused by magnetic moment rotation combined with the spin orbit coupling. This acceleration decrease the moment of inertia, since inertia is the tendency of staying in a constant magnetic state. Our analysis also shows that the moment of inertia is linked to the spin-polarization of the bands. Since, as mentioned, the inertia preserves the angular momentum, it has largest contributions  in the electronic structure, where multiple electron bands with the same spin-polarization are close to each other (cf. Fig.~\ref{fig4} c). However, some aspects of the inertia, e.g. being caused by band overlaps, is similar to the Gilbert damping \cite{Gilmore08}, although the moment of inertia is a property that spans over the whole band structure and not only over the Fermi-surface. Inertia is relevant in the equation of motion \cite{Gilmore07,Boettcher12c} only for $\tau\gtrsim\unit[0.1]{ps}$ and particularly for low dimensional systems. Nevertheless, in the literature there are measurements, as reported in Ref.~\onlinecite{Zhou10}, where the inertia effects are present.   

In summary, we have derived a theoretical model for the magnetic moment of inertia based on the torque-torque correlation model and provided first-principle properties of the moment of inertia that are compared to the Gilbert damping. The Gilbert damping and the moment of inertia are both proportional to the spin-orbit coupling, however, the basic electron band structure mechanisms for having inertia are shown to be different than those for the damping. We analyze details of the dispersion of electron energy states, and the features of a band structure that are important for having a sizable magnetic inertia. We also demonstrate that the torque correlation model provides identical  results to those obtained from a Greens functions formulation. Furthermore, we provide numerical values of the moment of inertia that are comparable with recent experimental measurements\cite{Li15b}. The calculated moment of inertia parameter can be included in atomistic spin-dynamics codes, giving a large step forward in describing ultrafast, sub-$\unit{ps}$ processes.

\paragraph*{Acknowledgements}
\label{sec:acknow}
The authors thank Jonas Fransson and Yi Li for fruitful discussions. The support of the Swedish Research Council (VR), eSSENCE and the KAW foundation (projects 2013.0020 and 2012.0031) are acknowledged. The computations were performed on resources provided by the Swedish National Infrastructure for Computing (SNIC).

\bibliographystyle{apsrev}
 \bibliography{thonig.bbl}

\end{document}